\documentclass[review,11pt]{elsarticle}
\makeatletter
\def\ps@pprintTitle{%
 \let\@oddhead\@empty
 \let\@evenhead\@empty
 \let\@oddfoot\@empty
 \let\@evenfoot\@empty
}
\makeatother
\usepackage{amsmath,amssymb}
\usepackage{amsthm}
\usepackage{graphicx,psfrag,epsf}
\usepackage{tikz-cd}
\usetikzlibrary{calc}
\usepackage{enumerate}
\usepackage{xcolor}
\usepackage{float}
\usepackage[colorlinks=true, linkcolor=blue, citecolor=blue, urlcolor=blue]{hyperref}
\usepackage{setspace}

\newtheorem{Lemma}{Lemma}
\newtheorem{Theorem}{Theorem}
\newtheorem{Corollary}{Corollary}
\newtheorem{Property}{Property}
\newtheorem{Example}{Example}
\newtheorem{Remark}{Remark}
\newtheorem{Assumption}{Assumption}

\usepackage[margin=1in]{geometry}

\begin{document}

\begin{frontmatter}

\title{Rational Expectations in Empirical Bayes
  \tnoteref{t1}
}

\tnotetext[t1]{
We thank Jiaying Gu for her generous and valuable comments.
}

\author[inst1]{Valentino Dardanoni}
\author[inst2]{Stefano Demichelis}
\address[inst1]{Department of Economics, Business and Statistics, University of Palermo}
\address[inst2]{Department of Economics and Management, University of Pavia}

\begingroup
\singlespacing
\begin{abstract}
We propose a principled framework for nonparametric empirical Bayes (EB) estimation, based on the idea that the prior should be consistent with the observed posterior and that Bayesian updating should be stable. Focusing on discretized priors, we characterize EB estimators as fixed points of a posterior belief operator. We establish the uniqueness of such fixed points and illustrate how the approach improves transparency and interpretability in standard EB settings, including a recent model of discrimination. 
\end{abstract}
\endgroup


\end{frontmatter}

\section{Introduction}
\label{sec:Introduction}

In a seminal paper, Robbins \cite{robbins1956} introduced the empirical Bayes (EB) framework: an unknown distribution $G$ generates a random sample $\alpha_1, \dots, \alpha_I$ of a latent parameter  $\alpha \in \mathbb{R}^d$: while the $\alpha_i$'s are unobserved, each produces an observed random variable $x_i$ independently via a known density $p(x_i \mid \alpha_i)$ for $i = 1, \dots, I$. Using the observed data $x$, researchers can infer properties of the compound distribution of the individual parameters $\alpha_i$. 
Like all Bayesian methods, EB estimation requires choosing a prior. The distinctive feature of EB methods -hence the term "empirical"- is that the prior is estimated directly from the data. This estimated prior is then used to compute the posterior for inference. In parametric EB, the prior $G$ is assumed to follow a specific parametric form, with its parameters estimated from the data. Nonparametric EB, by contrast, estimates $G$ flexibly, often using a discretized support.\footnote{For a detailed discussion of parametric and nonparametric approaches to modeling $G$, see \cite{Efron2010}, \cite{KoenkerGu2024}, \cite{IgnatiadisSen2024}, and the NBER presentation by \cite{gu2022}.}  

In many recent papers, researchers apply nonparametric maximum likelihood estimation  (NPMLE) to recover latent heterogeneity distributions without explicitly invoking the conceptual framework of priors and posteriors. While successful in practice, this usage often leaves the fundamental EB principle -updating a prior into a posterior based on observed data- somewhat implicit. In this paper, we propose a principled way to embed NPMLE within the spirit of EB by introducing two simple conditions that a rational forecaster might impose on priors: \emph{coherence} and \emph{stability}. Coherence requires that the prior coincide with the posterior: if beliefs are already updated by data, there is no reason to adjust them further. Stability gives the condition under which types given zero prior weight should not be revived by the data. These two properties mirror classic concepts from rational expectations and evolutionary game theory, and jointly define what we call Rational Expecations (RE) properties of empirical Bayesian updating.

We apply our RE approach to two distinct cases: one where observations are generated by a discrete density, and another where observations are continuous. In the discrete case, the map $F$ from the prior $\pi$ to the distribution of observations $f$,  $f=F\pi$, is generally not injective, leading to potential non-identifiability issues. However, we show that if the sample generated by $f$ has at least one configuration with zero probability (i.e., some point in the support of $f$ is unobserved in the data), we can establish uniqueness of $\pi$ (i.e., uniqueness of the discretized distribution of $G$). This follows from a simple geometric fact: although the map from the probability simplex of 
$\pi$ to $f$ is not injective in general, its restriction to the boundary is injective under a mild genericity condition. This is the key result of our paper, formalized in Theorem \ref{th:discr}.
\footnote{The theorem becomes particularly useful when the sample space is relatively large relative to the sample size, so that any sample we observe is likely to have no observations in some points of the sample space.}%
To illustrate the potential usefulness of the theorem, we apply Theorem \ref{th:discr} to discuss identification in the well known \citet{KlineWalters2021} discrimination model. 

Our approach provides both a justification for the use of NPMLE in empirical Bayes and a set of novel insights, particularly about uniqueness in overparameterized discrete settings, by reconnecting these methods to their Bayesian roots. In the rest of the paper, we extend Theorem \ref{th:discr} to the continuous setting in Theorem \ref{th:cont}, giving a simple and more general proof of earlier findings discussed in Lindsay \cite{lindsay1995}. Theorems \ref{th:conttop}  and \ref{th:asymptotical} briefly address the issue of the approximation of the true distribution $G$ by the discretized support. Proofs for all results are in the Online Appendix.


\section{Setting}
\label{sec:Setting}

We begin by assuming that the observable outcomes follow a discrete distribution, supported on a set $\mathcal{I}$ with cardinality $I$. Let $0 \leq b_i \leq I$ represent the number of units in the sample with the observed outcome $x_i \in \mathcal{I}$, which we collect in an $I$-dimensional vector $b$. Assume that there is a finite number $J$ of unobserved {\em types}, denoted as $t_j$ for $j=1,\dots,J$. In typical applications, each type corresponds to a specific combination of the parameters of interest.  Assuming observable outcomes follow a known discrete distribution  given the type, let $F_{ij} = \Pr(x_i \mid t_j)$, with  $0 < F_{ij} < 1$, and collect these probabilities in an $I \times J$ matrix $F$, whose elements are fixed by the statistical model and do not depend on the observed sample.

Given a prior distribution $\pi$ over types (a vector in the $J-1$ dimensional simplex $\Delta_{J-1}$), we apply Bayes theorem to compute the posterior probability that unit $i$ is of type $j$:
\[
\Pr(t_j \mid x_i) = \frac{ F_{ij} \pi_{j} }{ \sum_j F_{ij} \pi_{j}} = h_{ij}, \quad i=1,\dots,I; \ j=1,\dots,J.
\]
Collect $h_{ij}$ into the $I \times J$ matrix  $H$. Notice that $H$ is a function of the prior $\pi$  and the density matrix $F$, and that $b' H$ defines the posterior distribution, which we denote
\begin{equation}
\label{eq:bayes}
h_j(\pi) =  \frac{1}{\sum_l b_l} \sum_i b_i \frac{ F_{ij}}{\sum_k F_{ik} \pi_{k}} \pi_j,  \ j=1,\dots,J.
\end{equation}

The first challenge in any Bayesian approach is selecting an appropriate prior distribution.  
In this setting with discrete types, both $\pi$ and $h(\pi)$ are probability vectors defined on the same support.  
Our primary requirement is that the prior should not be contradicted by the inferences drawn from it. 
In other words, if the posterior (already informed by the data) matches the prior, then there is no reason to update the prior further. This is the notion of \emph{coherence}:
\begin{Property} 
\label{pr:coherence}
\hspace{-0.1cm} {\sc Coherence:} The prior distribution $\pi$ must match the posterior distribution, i.e.,  
$\pi = h(\pi)$.
\end{Property}
It is convenient to define a {\em discrepancy function} for each type $j$ as
\[
d_j(\pi) =  \sum_i \beta_i \left(\frac{ F_{ij}}{\sum_k F_{ik} \pi_k} -1\right), \ \ j = 1, \dots, J.
\]
$d_j(\pi)$ quantifies the average proportional excess (or deficit) of density mass assigned to type $j$ under the prior $\pi$. When $d_j(\pi) > 0$, data suggest type $j$ deserves more weigh; when $d_j(\pi) < 0$, the prior may be overemphasizing type $j$.

Under the coherence property, choosing an appropriate prior is equivalent to finding a fixed point of a system of nonlinear equations. Since $h$ is a continuous map from the probability simplex into itself, Brouwer’s fixed-point theorem guarantees the existence of at least one fixed point. The natural question is whether this fixed point is unique. The answer, however, is clearly negative: any vertex of the simplex is a fixed point.
If the prior assigns probability one to a single type, the posterior must coincide with the prior, as there is no probability mass to shift elsewhere:
the faces of the simplex are invariant under the mapping $h$. On the other hand, we cannot require full support of $\pi$ as we expect many zero elements when  data is limited relative to parameter configurations.

To gain insight into when zero priors may be rationally expected and when they may not, consider the following example:
\begin{Example}
Suppose there are two unobserved types of goalkeepers (GKs): `bad', with a probability of saving a penalty equal to 0.05, and `good', with a probability equal to 0.9. Suppose we observe two GKs and a single penalty shot: one saves the shot, and the other does not, so $ b_1 = b_2 = 1$. In this situation, coherent Bayesian updating yields three fixed points: $\pi^* \in \{ (1, 0),\ (0.471, 0.529),\ (0, 1) \}.$ A rational predictor might regard the interior fixed point as the most plausible: when $\pi_j$ is small, updating increases it (i.e., $d_j(\pi) > 0$) and the type is ``revived'' (the data support its inclusion). Suppose instead that $b_1 = 4$ and $b_2 = 96$. In this case, the hypothesis that all GKs are of the good type and that four were merely unlucky becomes plausible. The data overwhelmingly support the good type, and the posterior operator $h(\pi)$ pulls all priors toward $\pi_2 = 1$. When $\pi_1$ is near zero, updating further lowers it: $d_1(\pi) < 0$, meaning that type 1 remains ``dead.''  This contrasts with the earlier case, where both types were empirically supported and $d_j(\pi) > 0$ near zero for both types. $\triangle$
\end{Example}
In general, a rational forecaster would assign zero probability to a certain type only if, during the process of updating priors near that point, the probability assigned to that type decreases.
\begin{Property}
\label{pr:stability}
\hspace{-.1cm} {\sc Stability:} A fixed point \( \bar{\pi} \) of the mapping \( h \) is said to be \emph{stable} if, whenever \( \bar{\pi}_j = 0 \), it holds that \( d_j(\bar{\pi}) \leq 0 \).
\end{Property}
\emph{Stability} parallels the concept of an Evolutionarily Stable Strategy (states that, once established, cannot be invaded by rare, disadvantaged alternatives) or, more vividly, serves as a ``no zombie'' condition: types that are eliminated at any point do not get revived by posterior updating. \emph{Coherence}, on the other hand, mirrors the concept of a Rational Expectations Equilibrium: agents’ beliefs about the environment match the empirical distribution generated by their behavior. Together, Coherence and Stability express a general Rational Expectations principle: priors should be self-confirming and robust to local empirical evidence.

Importantly, both properties also emerge naturally from nonparametric likelihood maximization. In this setting, the marginal log-likelihood is given by
\[
L(\pi) = \sum_{i} b_i \log\left( \sum_{j} F_{ij} \pi_j \right),
\]
and it can be readily verified that the Kuhn–Tucker  conditions for maximizing $\L(\pi)$ over the probability simplex imply the following:
\begin{itemize}
    \item When \( \pi_j > 0 \), the condition \( h_j(\pi) = \pi_j \) must hold (Coherence)
    \item When \( \pi_j = 0 \), it must be that \( d_j(\pi) \leq 0 \) (Stability)
\end{itemize}
We will appeal to this fact in the proof of Lemma 1 below.


\section{Results}
\label{sec:Results}

\subsection{Discrete Case}
\label{subsec:DiscreteCase}

In this section, we establish the conditions under which Properties~\ref{pr:coherence} and~\ref{pr:stability} jointly ensure the uniqueness of the fixed point. Before stating the main results, we introduce some notation and define the relevant spaces and functions.
Let $Q_1, \dots, Q_I$ be the vertices of the $(I-1)$-dimensional simplex $\Delta_{I-1}$. We denote generic vectors in this simplex by $\tau$. Define $\beta_i = \frac{b_i}{\sum_{k=1}^I b_k}$ as the fraction of observations with value $i = 1, \dots, I$, so that   $\beta = (\beta_1, \dots, \beta_I)$ is a vector in $\Delta_{I-1}$. The matrix $F = (F_{ij})$ defines a map $F: \pi \in \Delta_{J-1} \mapsto \tau(\pi) \in \Delta_{I-1}$, given by $\tau_i = \sum_j F_{ij} \pi_j$. The $j$-th column of $F$ will be denoted by $F_j$.

To handle cases where some $\beta_i = 0$, we collapse all vertices of $\Delta_{I-1}$ corresponding to $\beta_i = 0$ into a single point $O$, forming an auxiliary space $\Delta^0_{\bar{I}}$. Formally, if $\bar{\mathcal{I}} = \{ i \mid b_i \neq 0 \} \quad \text{and} \quad \bar{I} = |\bar{\mathcal{I}}|$, we define the index set $\mathcal{I}^0 = \bar{\mathcal{I}} \cup \{0\}$, and the collapsed simplex as
\[
\Delta^0_{\bar{I}} = \left\{ \tau_i \;\middle|\; i \in \mathcal{I}^0,\ \tau_i \geq 0,\ \sum_{i \in \mathcal{I}^0} \tau_i = 1 \right\}.
\]

By composing $F$ with the collapsing map, we obtain the affine map $F^0: \Delta_{J-1} \rightarrow \Delta^0_{\bar{I}}$, whose associated $(\bar{I}+1) \times J$ matrix is:
\[
\left\{ F_{0j} = 1 - \sum_{i \in \bar{\mathcal{I}}} F_{ij},\quad F_{ij} \;\middle|\; i \in \bar{\mathcal{I}},\ j = 1, \dots, J \right\}.
\]

\begin{figure}[htbp]
  \centering
  \includegraphics[width=0.7\textwidth]{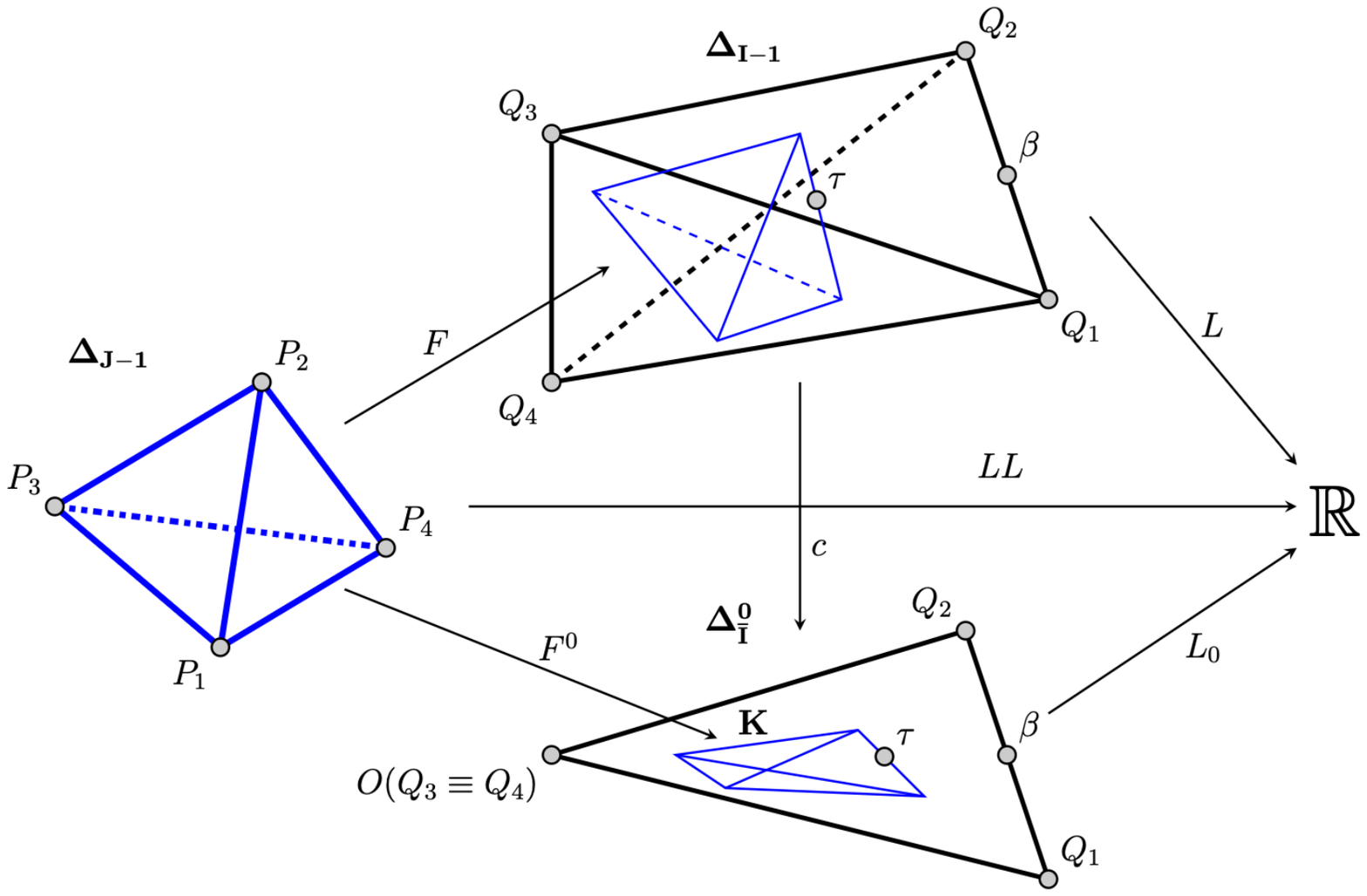} 
  \caption{Some maps and spaces.}
  \label{fig:fattori}
\end{figure}

Define the function $L(\tau) = \sum_{i \in \mathcal{I}} \beta_i \ln \tau_i$. When all $\beta_i \neq 0$, it is well-defined and strictly convex on $\Delta_{I-1}$. If some $\beta_i = 0$, it remains well-defined and strictly convex on the collapsed simplex $\Delta^0_{\bar{I}}$. By composing $L$ with $F$ or $F^0$, we obtain the log-likelihood function:
$L(\pi) = \sum_{i \in \mathcal{I}} \beta_i \ln \tau_i(\pi)$. See Figure \ref{fig:fattori} for an illustration of the relevant maps and spaces.

We start with the following:
\begin{Lemma}
\label{lm:Kull}
$\bar{\pi}$ is a stable fixed point if and only if it minimizes the Kullback–Leibler divergence:
\[
D_{KL}(\beta \,\|\, \tau(\bar\pi)) = \sum_{i \in \mathcal{I}} \beta_i \ln \frac{\beta_i}{\tau_i(\bar \pi)}.
\]
\end{Lemma}
\begin{Remark}
\label{rm:ML}
In this context, minimizing $D_{KL}$ is equivalent to maximizing the log-likelihood,
since they differ only by a sign and the constant term $\sum_{i \in \mathcal{I}} \beta_i \ln \beta_i$.
\end{Remark}
When no risk of confusion arises, we will treat $L$ as a function of either $\tau$ or $\pi$, depending on the context. This lemma will be useful in two ways. On the one side, it provides a bridge with the methods of likelihood maximization giving them a solid EB conceptual foundation, on the other it provides a useful technical tool insofar we will use the function $D_{KL}$, or $L(\pi)$ as a, loose, analogue of a Liapunov function whose discretized gradient flow is given by \ref{eq:bayes}.%
\footnote{Key references to nonparametric maximum likelihood estimation are
\cite{kiefer1956},  \cite{laird1978}, and \cite{KoenkerMizera2014}.}

The assumption that $F_{ij} > 0$ is, in general, not sufficient to ensure the uniqueness of $\bar{\pi}$, as the following example illustrates.
\begin{Example}
\label{ex:GK2}
Suppose we want to evaluate the ability of a group of eight goalkeepers, each facing two penalty shots. Three GKs fail to save any penalties, three save one, and two save both, $\beta = \left( \frac{3}{8}, \frac{3}{8}, \frac{2}{8} \right)$.
Let the support of $\pi$ be $(0.1,, 0.2, \dots, 0.9)$, so $J = 9$. Clearly, multiple solutions $\bar{\pi}$ satisfy both the Coherence and Stability conditions while exactly matching the observed data.%
This is unsurprising: we are attempting to estimate 9 parameters to match only 3 empirical moments. $\triangle$
\end{Example}
Are there conditions that imply uniqueness of the fixed point $\bar{\pi}$? Uniqueness follows from the conjunction of the following two facts:
\begin{enumerate}
\item When all $b_i \neq 0$, the maximizer $\tau$ of $L(\tau)$ is unique on $\Delta_{I-1}$; if some $b_i = 0$, it is unique on $\Delta^0_{\bar{I}}$.
\item The preimage $F^{-1}(\tau)$ (or $(F^0)^{-1}(\tau)$, respectively) in $\Delta_{J-1}$ is unique.
\end{enumerate}
Clearly, strict concavity of the likelihood function ensures point (1), and point (2) holds generically, which is typical when $J \leq I$. The key insight is that uniqueness can also arise in overparameterized settings (i.e., $J > I$) when the data are incompatible with the image of $F$. Suppose that $\beta$ lies outside the image of $F$; this occurs, for example, when some $b_i = 0$, because we have assumed that all $F_{ij} > 0$, and so $F^0(\Delta_{J-1})$ does not touch the boundary. In this case, it is easy to see that the $\tau$ in $\Delta_{I-1}$ (or in $\Delta^0_{\bar{I}}$ if some $b_i = 0$) that maximizes the function $L$ must lie on the boundary of $F(\Delta_{J-1})$ or $F^0(\Delta_{J-1})$, and will be unique, as stated above.

Now, it is an elementary fact, proven in the Appendix, that an affine map from one polyhedron to another (here, from $\Delta_{J-1}$ to $\Delta_{I-1}$ or $\Delta^0_{\bar{I}}$) is injective on the preimage of the boundary, regardless of the size of $J$, provided a certain genericity condition (spelled out below) is satisfied. Thus, injectivity holds in this case as well. This is illustrated in the next example:
\begin{Example}
\label{ex:GK2a}
Suppose 8 GKs, now facing 10 penalties each, and $\beta = \frac{1}{8}(1, 0, 0, 0, 2, 3, 1, 0, 0, 0, 1).$ Let the support of $\pi$ consist of $J = 999$ points in $(0.001, 0.999)$. No $\pi \in \Delta^{J-1}$ satisfying coherence and stability can exactly match $\beta$ (i.e., there is no $\pi$ such that $\beta = F\pi$), due to the presence of zeros in $\beta$. This imposes additional constraints and eliminates many potential values in the support of $\pi$. The solution for $\pi$ is unique, and its support consists of just three non-zero points: $\pi(.001)=0.1240$, $\pi(.483)=0.7516$, $\pi(0.9)=0.1244$.
$\triangle$
\end{Example}
If we denote by $\bar{F}_j = \big(1, F_{ij} \mid i \in \bar{\mathcal{I}} \big)$ the column vector of size $\bar{I}+1$, the following assumption ensures injectivity:
\begin{Assumption}
\label{as:gendiscr}
If $\bar{\mathcal{I}} = \mathcal{I}$, then the matrix $(F_1, \dots, F_J)$ has rank $J$. If $\bar{\mathcal{I}} \subset \mathcal{I}$, then for any subset $S = \{\bar{F}{j_1}, \dots, \bar{F}{j_s} \}$, with $s = \min(J, \bar{I} + 1)$, the matrix $(\bar{F}{j_1}, \dots, \bar{F}{j_s})$ has rank $s$.
\end{Assumption}
We refer to the online Appendix for the geometrical ibterpretation of this assumption.
\begin{Remark}
This assumption requires a kind of “restricted injectivity” on the relevant subsets when $\bar{\mathcal{I}} \subset \mathcal{I}$. Specifically: 1) If $J \leq \bar{I}$, we require that $(\bar{F}_1, \dots, \bar{F}_J)$ is injective;
2) If $J \geq \bar{I} + 1$, it suffices that for any choice of $\bar{I} + 1$ vectors $(\bar{F}{j_1}, \dots, \bar{F}{j_{\bar{I}+1}})$ we have $\det\left(\bar{F}{j_1}, \dots, \bar{F}{j_{\bar{I}+1}}\right) \neq 0$.\footnote{For a more geometric interpretation, see also Assumption 1 in the Online Appendix.}
\end{Remark}
Note that Assumption~\ref{as:gendiscr} is generic, in the sense that for any matrix, there always exists an arbitrarily small perturbation of its coefficients that ensures the assumption holds. Moreover, if the assumption holds at a given point, it also holds throughout a neighborhood of that point.

The assumption could in fact be weakened. As will be evident from the proof in the next section, it suffices to assume, for instance, that $\beta \notin F(\Delta_{J-1})$, and that any set of vectors $(\bar{F}{j_1}, \dots, \bar{F}{j_s})$, whose convex hull lies on the boundary, is linearly independent. This latter condition will be the key requirement in the proof of Theorem~\ref{th:conttop}, and can also be used in practice to reduce computational burden.
\begin{Theorem}
\label{th:discr}
Suppose that Assumption~\ref{as:gendiscr} holds. Then there exists a unique stable fixed point $\bar\pi$. Moreover, if $\bar{\mathcal{I}} \subset \mathcal{I}$, the support of $\bar\pi$ consists of at most $\bar{I}$ points.
\end{Theorem}
The proof is in the Appendix. The key insight of this theorem is that an apparent lack of information (specifically, when some $b_i$ are equal to zero) ensures the uniqueness of the posterior distribution.\footnote{This echoes a classical result in the moment problem: when the vector of moments lies on the boundary of the moment space, the representing measure is unique.}  Intuitively, this introduces additional constraints: the solution must lie on the boundary of the probability simplex, effectively reducing the number of unknown parameters. The bound on the support of $\bar{\pi}$ is also useful in applications, leading to significant computational savings. After a few iterations, the weights of most $j$'s converge to zero and can be discarded, allowing computations to focus on the relatively few nonzero terms. The next example applies the theorem to real data.  
\begin{Example}
\citet{KlineWalters2021} study discrimination in the labor market using data from correspondence experiments, where researchers send out $L$ fictitious job applications from each of two groups, labeled $a$ and $b$, and record the number of callbacks for each group. For each job $i = 1, \dots, n$, the data are summarized by $Z_i = (C_{ai}, C_{bi})$, where ($C_{ai}$, $C_{bi}$) denote the number of callbacks received by groups $a$ and $b$. The density is specified as independent bivariate binomial:
\begin{equation*}
    p(z \mid \theta) = \binom{L}{c_a} \binom{L}{c_b} p_a^{c_a} (1 - p_a)^{L - c_a} p_b^{c_b} (1 - p_b)^{L - c_b},
\end{equation*}
where $z = (c_a, c_b)$ and $\theta = (p_a, p_b)$ represent the group-specific callback probabilities. The objective is to estimate the posterior probability that a job with observed callback pattern $z$ discriminates against group $b$,
$\theta_{\text{discr}}(G; z) := \mathbb{P}_G[p_a > p_b \mid Z = z]$, where $G$ denotes the structural distribution of parameters $(p_a, p_b)$ across jobs. 

\citet{KlineWalters2021} apply EB estimation to gender discrimination data by \citet{ArceoCampos2014}. With $L=4$, the data consist of the frequencies for the $5 \times 5$ callback patterns, observed across 799 jobs. These frequencies are reported in Table~\ref{tab:discrimination}, with subscripts $f$ and $m$ denoting female and male applicants, respectively. \citet{GuIgnatiadisShaikh2025} revisits the estimation of $\theta_{\text{discr}}(G; z)$, highlighting the challenges reflected in the non-uniqueness of the NPMLE. At first glance, this non-uniqueness may appear surprising: a naive application of Theorem~\ref{th:discr} would suggest uniqueness, since the empirical frequency vector $b$ (reported in column 1 of Table~\ref{tab:discrimination}) contains a zero entry -specifically, for the callback pattern $(1,4)$. 

This apparent inconsistency with Theorem 1 is resolved by recognizing that the independence assumption between the two binomial callback distributions imposes very strong structural restrictions on the density matrix $F$. While this assumption simplifies the model, it is arguably unrealistic in the context of discrimination detection, and it introduces degeneracies that obstruct identification
 (we reject the null of independence using a simulated Fisher exact test, with a $p$-value below $10^{-4}$). 
 
 Several well-known parametric alternatives relax the independence assumption by allowing dependence between the binomial marginals, such as the copula-based one of \citet{PanagiotelisCzadoJoe2012}. While more flexible than the independent model, they still impose low-dimensional, non-generic restrictions on the joint distribution of callbacks. An alternative, nonparametric approach is to sample joint discrete distributions within a Fréchet class. Given marginal binomial distributions with parameters $p_f$ and $p_m$, we sample $k$ joint distributions with the corresponding marginals. The set of feasible tables corresponds to a convex polytope defined by linear equality and inequality constraints. To sample approximately uniformly from this set, we use an MCMC algorithm based on \emph{2-by-2 switch} moves (also known as \emph{Markov basis moves}; see \citealp{DiaconisSturmfels1998}). At each step, two distinct rows $i, j$ and columns $k, l$ are selected, and a switch is performed on the $2 \times 2$ submatrix, preserving row and column sums. This guarantees that each sampled table matches the specified marginals exactly, ranges over the full Fréchet class, and is drawn from an approximately uniform distribution.
 
We discretize $p_f$ and $p_m$ into 99-point grids on $(0.01, 0.99)$, and for each of the resulting $99^2$ marginal pairs, we draw 99 MCMC samples. Thus, $F$ has dimension $I = 25$ and $J = 99^3$. The fixed point $\hat{\pi}$ has 24 nonzero  (i.e. below $10^{-6}$) entries. We compute the estimated conditional callback and  discrimination probabilities, reported in Table~\ref{tab:discrimination}. Notably, the estimated distribution of callbacks $p$ is strikingly close to the observed one $\beta$.

 Looking at discrimination for specific callback patterns, \citet{KlineWalters2021} estimate discrimination probabilities of 0.72 and 0.99 for the callback patterns $(1,0)$ and $(4,0)$ under the independent binomial model. Our corresponding estimates are 0.41 and 0.97. \citet{GuIgnatiadisShaikh2025}, accounting for non-uniqueness and sampling variation, report lower bounds of the 95\% confidence interval below 0.02 for $(1,0)$ and around 0.9 for $(4,0)$.
This example suggests that a major source of the large sampling variation in the estimates of \citet{KlineWalters2021}, as discussed in \citet{GuIgnatiadisShaikh2025}, lies in the non-generic restrictions imposed by the independence assumption on the joint distribution of callbacks. Our Theorem~\ref{th:discr} clarifies that, and shows that, when a more refined callback model is used the difficulty is circumvented. We think that this ability to make the shortcomings of a model emerge is one of the strong points of our method. $\triangle$
\vspace{-.4cm}
\begin{footnotesize}
\begin{table}[H]
\centering
\caption{Data and estimates by callback pattern}
\label{tab:discrimination}
\begin{tabular}{ccccccccc}
\hline
Callback & Count & $\beta$ & $p$ & $p_f$ & $p_m$ & Pr($p_f > p_m$) & Pr($p_f = p_m$) & Pr($p_f < p_m$) \\
\hline
(0,0) & 573 & 0.7171 & 0.7161 & 0.0126 & 0.0123 & 0.0038 & 0.9903 & 0.0059 \\
(0,1) & 26  & 0.0325 & 0.0328 & 0.1326 & 0.0164 & 0.0112 & 0.6815 & 0.3073 \\
(0,2) & 11  & 0.0138 & 0.0135 & 0.4012 & 0.1340 & 0.1452 & 0.0294 & 0.8254 \\
(0,3) & 3   & 0.0038 & 0.0038 & 0.4177 & 0.0924 & 0.0600 & 0.0001 & 0.9399 \\
(0,4) & 2   & 0.0025 & 0.0025 & 0.7448 & 0.1190 & 0.0807 & 0.0000 & 0.9193 \\
(1,0) & 30  & 0.0375 & 0.0386 & 0.0605 & 0.2262 & 0.3976 & 0.5893 & 0.0131 \\
(1,1) & 10  & 0.0125 & 0.0130 & 0.2225 & 0.2632 & 0.3462 & 0.5233 & 0.1305 \\
(1,2) & 8   & 0.0100 & 0.0102 & 0.4887 & 0.5358 & 0.4568 & 0.0004 & 0.5428 \\
(1,3) & 4   & 0.0050 & 0.0053 & 0.5669 & 0.4271 & 0.2196 & 0.0003 & 0.7802 \\
(1,4) & 0   & 0 & 0.0000 & 0.4677 & 0.5103 & 0.6910 & 0.0074 & 0.3016 \\
(2,0) & 16  & 0.0200 & 0.0194 & 0.0505 & 0.5348 & 0.9855 & 0.0014 & 0.0131 \\
(2,1) & 12  & 0.0150 & 0.0147 & 0.3706 & 0.5262 & 0.8954 & 0.0255 & 0.0791 \\
(2,2) & 12  & 0.0150 & 0.0144 & 0.4691 & 0.5406 & 0.5843 & 0.0033 & 0.4124 \\
(2,3) & 7   & 0.0088 & 0.0086 & 0.5005 & 0.5405 & 0.3643 & 0.0001 & 0.6356 \\
(2,4) & 6   & 0.0075 & 0.0072 & 0.6298 & 0.5431 & 0.1991 & 0.0016 & 0.7994 \\
(3,0) & 13  & 0.0163 & 0.0166 & 0.0664 & 0.5336 & 0.9715 & 0.0000 & 0.0285 \\
(3,1) & 10  & 0.0125 & 0.0129 & 0.4546 & 0.5672 & 0.5453 & 0.0001 & 0.4546 \\
(3,2) & 7   & 0.0088 & 0.0088 & 0.4409 & 0.5495 & 0.8700 & 0.0010 & 0.1289 \\
(3,3) & 8   & 0.0100 & 0.0104 & 0.4888 & 0.5360 & 0.4960 & 0.0014 & 0.5027 \\
(3,4) & 2   & 0.0025 & 0.0026 & 0.9152 & 0.6852 & 0.0625 & 0.2991 & 0.6384 \\
(4,0) & 11  & 0.0138 & 0.0138 & 0.0460 & 0.8314 & 0.9788 & 0.0000 & 0.0212 \\
(4,1) & 2   & 0.0025 & 0.0025 & 0.4733 & 0.7672 & 0.8417 & 0.0000 & 0.1582 \\
(4,2) & 6   & 0.0075 & 0.0074 & 0.4895 & 0.6819 & 0.6341 & 0.0004 & 0.3655 \\
(4,3) & 4   & 0.0050 & 0.0051 & 0.5444 & 0.7560 & 0.7879 & 0.1529 & 0.0593 \\
(4,4) & 16  & 0.0200 & 0.0200 & 0.9670 & 0.9703 & 0.0410 & 0.9358 & 0.0232 \\
\hline
\end{tabular}
\end{table}
\end{footnotesize}
\end{Example} 


\subsection{Continuous Case}
\label{subsec:ContinuousCase}

In many applications, it is more natural to assume that the observed random variable is continuous. This motivates the following variant of the theorem. Suppose that the observations $x_1, \dots, x_n, \dots, x_N$ come from a continuous distribution, and let the density of $x_n$ conditional on being of type $t_j$ be known and equal to $F_{nj} = p(x_n \mid t_j)$, a continuous function in $x_n$.

The relevant assumption now takes the following form:
\begin{Assumption}
\label{as:gencont}
Let $\bar{F}_j = \big( 1, F_{nj} \mid n = 1, \dots, N \big)$, a column vector of size $N + 1$. Then, for any choice of $N + 1$ such vectors 
$(\bar{F}_{j_1}, \dots, \bar{F}_{j_{N + 1}})$, we have
$\det \big( \bar{F}_{j_1}, \dots, \bar{F}_{j_{N + 1}} \big) \neq 0$.
\end{Assumption}
The theorem becomes:
\begin{Theorem}
\label{th:cont}
Suppose there exists an $x \notin \{x_1, \dots, x_N\}$ such that $p(x \mid t_j) \neq 0$ for all $j$. If Assumption~\ref{as:gencont} is satisfied, then the stable fixed point is unique.
\end{Theorem}
The proof is in the Appendix. In the case where the statistical model belongs to the exponential family, the upper bound on the number of nonzero support points in the fixed point has been recently sharpened by \citet{PolyanskiyWu2020}.


\subsection{$\pi$ and $g$}
\label{subsec:pi_and_g}

So far, we have assumed that the set of $J$ types (the support of $\pi$) is finite and given. In many cases, however, it is more natural to assume that $j$ is a continuous variable. We provide a version of our theorem in this setting as well. Moreover, since computations can only be performed after discretizing the type space, it is important to verify that, as the discretization grid becomes finer, the fixed point over the discrete set approximates the true one in the continuum. This is the content of Theorem~\ref{th:conttop} below.

A key point is that the bound on the size of the support of the fixed point is uniform across discretizations and carries over to the continuum. Before stating our results, we introduce some additional notation. We denote the unobserved types by $t$, assumed to be continuous and to lie in a compact topological space $T$—for example, a compact subset of $\mathbb{R}^n$. We discretize $T$ using increasingly finer grids: let $T_n$ be an increasing sequence of finite subsets of $T$, such that $T_n \subset T_{n+1} \subset \dots$ and $\bigcup_n T_n$ is dense in $T$. Let $\mathcal{P}(T)$ and $\mathcal{P}(T_n)$ denote the spaces of probability measures on $T$ and $T_n$, respectively, with the natural inclusions $\mathcal{P}(T_n) \subset \mathcal{P}(T_{n+1}) \subset \dots \subset \mathcal{P}(T)$. Probability measures will be denoted by $d\pi(t)$ for $T$ and $d\pi_n(t)$ for $T_n$.

Suppose there is a finite set of outcomes $x_i$, indexed by $i \in I$, and that the probability of outcome $x_i$ conditional on $t$ is given by $p(x_i \mid t)$. There is a natural map $F: \mathcal{P}(T) \to \Delta_{I-1}$, where $\Delta_{I-1}$ is the $(I-1)$-dimensional simplex, given by:
\begin{equation}
\label{eq:maptop}
F(d\pi(t))_i = \int p(x_i \mid t) \, d\pi(t)
\end{equation}
The continuous analogue of equation~\ref{eq:bayes} becomes:
\begin{equation}
\label{eq:bayestop}
h(d\pi(t)) = \left(\frac{1}{\sum_l b_l} \sum_i b_i \frac{p(x_i \mid t)}{\int p(x_i \mid t) \, d\pi(t)} \right) d\pi(t)
\end{equation}
The continuous analogue of Property~\ref{pr:coherence} is:
\begin{equation}
\label{eq:coherenttop}
\frac{1}{\sum_l b_l} \sum_i b_i \left( \frac{p(x_i \mid t)}{\int p(x_i \mid t) \, d\pi(t)} - 1 \right) = 0
\end{equation}
The continuous analogue of Property~\ref{pr:stability} is:
\begin{equation}
\label{eq:stabletop}
\frac{1}{\sum_l b_l} \sum_i b_i \left( \frac{p(x_i \mid t)}{\int p(x_i \mid t) \, d\pi(t)} - 1 \right) \leq 0
\end{equation}
For any $n$, there exists at least one measure, denoted by $d\bar\pi_n(t)$, satisfying 
Property~\ref{pr:coherence} and Property~\ref{pr:stability}. A measure satisfying equations~\ref{eq:coherenttop} and~\ref{eq:stabletop} will be denoted by $d\bar\pi(t)$.

Although the map $F$ takes an infinite-dimensional space to a finite-dimensional one and is not injective, it still has an inverse on the boundary of the image under mild conditions, which we now spell out. Let $K = \text{co}(F(T)) = F(\mathcal{P}(T)) \subset \Delta_{I-1}$ be the convex hull of $H = F(T) \subset \Delta_{I-1}$. Denote the boundary of $K$ by $\partial K$, and let $\bar{I}$ be as before.
\begin{Assumption} \label{as:gentop}
For any set of $\bar{I}+1$ elements $t_1, \dots, t_{\bar{I}+1}$ such that $\text{co}(F(t_1), \dots, F(t_{\bar{I}+1})) \subset \partial K$, the points $F(t_1), \dots, F(t_{\bar{I}+1})$ are affinely independent.
\end{Assumption}
%
%
\begin{Theorem}
\label{th:conttop}
If Assumption~\ref{as:gentop} holds, there exists a unique measure $d\bar\pi(t)$ on $T$ satisfying equations~\ref{eq:coherenttop} and~\ref{eq:stabletop}. This measure maximizes the log-likelihood: for any other measure $d\pi$, we have  
\begin{small}
\[
L\left(F(d \bar\pi)\right) = \frac{1}{b} \sum_i b_i \, \ln \left( \int p(x_i \mid t) \, d\bar\pi(t) \right) \geq \frac{1}{b} \sum_i b_i \, \ln \left( \int p(x_i \mid t) \, d\pi(t) \right) = L\left(F(d\pi)\right).
\]
\end{small}
Moreover, any sequence of measures $\{d\bar\pi_n(t)\}$ satisfying Property~\ref{pr:coherence} and Property~\ref{pr:stability} on $T_n$ converges to $d\bar\pi(t)$ in the weak* topology. All such measures have support on at most $\bar{I}$ points.
\end{Theorem}

The proof is in the Appendix. Suppose now that the observations $x_1, x_2, \dots, x_k$ take values in a compact domain $X \subset \mathbb{R}^m$, and that their number increases. Under the hypothesis in Assumption~\ref{as:gentop}, for each $k$ there exists a unique fixed point $\bar\pi_k(t) \in \mathcal{P}(T)$ satisfying equations~\ref{eq:coherenttop} and~\ref{eq:stabletop}. On the other hand, the Glivenko–Cantelli theorem implies that the empirical measure associated with the observations converges to the true probability distribution of $x$ on $X$, which we denote by $d\alpha(x)$.

This raises some natural questions: What can be said about the measures $d\bar\pi_k(t)$ for large $k$? What are their limit points? What is their relationship to the distribution $d\alpha$? To be more precise, let the Lebesgue measures on $T$ and $X$ be denoted by $dt$ and $dx$, respectively, and assume that $d\alpha$ is absolutely continuous with respect to $dx$, i.e., $d\alpha(x) = a(x) \, dx$. The map in equation~\ref{eq:maptop} becomes 
\[
F(d\pi(t)) = d\tau(x) = \left(\int p(x \mid t) \, d\pi(t) \right) dx.
\]
To avoid uninteresting edge cases, we assume that both $a(x)$ and $p(x \mid t)$ are continuous and strictly positive, so that the Radon–Nikodym derivative $F(d\bar\pi)/d\alpha$ is well-defined, continuous, and nonzero. We then have the following result:
\vspace{-.1cm}
\begin{Theorem}
\label{th:asymptotical}
Let $k \to \infty$, and let $d\bar\pi(t)$ be any limit point of the sequence $\{d\bar\pi_k(t)\}$ in the weak* topology (such a point exists because $\mathcal{P}(T)$ is weak* compact). Then $d\bar\pi(t)$ minimizes the Kullback–Leibler divergence 
\[
D_{KL}(d\alpha \,\|\, F(d\bar\pi)) = \int \log \left( \frac{a(x)}{\int p(x \mid t) \, d\bar\pi(t)} \right) a(x) \, dx
\]
almost surely.
\end{Theorem}
The proof is in the Appendix. In general, $F(\mathcal{P}(T))$ is a strict subset of $\mathcal{P}(X)$, so there is no reason to expect that $F(d\bar\pi)$ coincides with $d\alpha$. However, when $d\alpha \in F(\mathcal{P}(T))$, i.e., when $a(x) = \int p(x \mid t) g(t) \, dt$ for some continuous $g(t)$, we obtain:
\begin{Corollary}
\label{cor:asymptotical}
If $d\alpha = F(d\pi^*)$ for some $\pi^* \in \mathcal{P}(T)$, and $F$ is injective, then $d\bar\pi_k(t)$ converges weakly* to $g(t) \, dt$.
\end{Corollary}
This follows immediately from the theorem, since the Kullback–Leibler divergence attains its minimum value (zero), and injectivity of $F$ ensures uniqueness. Note that $d\bar\pi_k(t)$ are discrete measures with finite support, while $g(t) \, dt$ is absolutely continuous with respect to $dt$, so weak* convergence is the strongest type of convergence that can be expected.

\newpage

\vspace{-.5cm}
     
     \begingroup
\singlespacing
\begingroup
\singlespacing
\bibliographystyle{chicago}
\bibliography{EB.bib}
\endgroup  
\endgroup

\newpage

\section{Appendix: Proofs}
\label{sec:Appendix}

Recall that:
\begin{itemize}
\item $\Delta_{J-1}$ is a simplex with generic element $\pi= (\pi_1,\, \dots, \, \pi_J)$,
\item $\Delta_{I-1}$ is a simplex with generic element $\tau = (\tau_1,\, \dots, \, \tau_I)$,
\item $ \bar {\mathcal{I}} =\{i \;  | \; b_{i} \neq 0\}$.
\end{itemize}
If  $\bar {\mathcal{I}}\subset {\mathcal{I}}$  properly, we let  $ {\mathcal{I}}^0  =\{0\} \bigcup  \bar {\mathcal{I}}$. Let us define the simplexes:
\begin{itemize}
\item $\bar \Delta_{\bar I -1} =\{ \tau_{i} \, | \, i \in \bar {\mathcal{I}}, \tau_i \geq 0, \sum_{i \in \bar I} \tau_{i}  =1\}$, 
\item  $\Delta^0_{\bar I} =  \{  \tau_{i} \, | \, i \in  \mathcal{I^0} , \tau_{i}  \geq 0, \sum_{i \in {\mathcal{I}}^0} \tau_{i} =1 \}$.
\end{itemize}
Note that $\bar\Delta_{\bar I -1}$ is in a natural way a face of  $\Delta^0_{\bar I }$, namely it consists of points $(0, \tau_1, \dots ,\tau_{\bar I})$. A generic point $(\tau_0, \, \tau_1, \dots, \tau_{\bar I}) \in \Delta^0_{\bar I } $  can always be written as  $ \mathbf{\tau}=((1-t), t\tau'_1, \dots, t\tau'_{\bar I})$ with $0\leq t \leq 1$ and $(\tau'_1, \dots, \tau'_{\bar I}) \in \bar\Delta(q)$.
Note also that  $\Delta^0_{\bar I} $  is obtained from $\Delta_{I-1}$ by collapsing all the vertices $Q_i, i|b_i=0$ to the  zero vertex.
The collapsing map $c$ is : 
\[
c: \: \tau_0=\sum_{i \notin  \bar {\mathcal{I}}} \tau_{i}= 1-\sum_{i \in  \bar {\mathcal{I}}} \tau_{i}
\]
and leaves the $\tau_{i},\, i \in \bar {\mathcal{I}}$ unchanged, in other words the following diagram commutes:
\[
 \begin{tikzcd}
 \bar\Delta_{\bar I -1} \arrow[rd, hook, "i"] \arrow[r,hook, "i"] & \Delta_{I-1} \arrow[d,two heads, "c"]\\
& \Delta^0
\end{tikzcd}
\]
where $i$ is the inclusion and $c$ is the collapsing map. Here and in the sequel, the hook before the arrow denotes an injective map, the double arrow after it a surjective one.

We define a new matrix:
$  F^0 = \{F_{0j}=1-\sum_{i \in  \bar{\mathcal{I}}} F_{ij}, \, F_{ij}  \, | \, \; i \in  \bar {\mathcal{I}} , j= 1,\, \dots,\, J \}$  and the corresponding  map  $\Delta_{J-1} \xrightarrow{F^0} \Delta^0_{\bar I}$ such that the diagram commutes:
\[
 \begin{tikzcd}
 \Delta_{J-1} \arrow[rd, "F^0"] \arrow[r, "F"] & \Delta_{I-1} \arrow[d,two heads, "c"]\\
& \Delta^0_{\bar I}
\end{tikzcd}
\]

If $P_1, P_2, \dots, P_J$  are the vertices  of  $\Delta_{J-1}$ , $F^0(\Delta_{J-1})$  will be the convex hull of the $ F^0(P_1), \dots, F^0(P_J)$ , we will call it $K$. Generic points $\tau$  and $ F^0(P_1), \dots, F^0(P_J)$ will be thought of as column vectors. If $\mathbf{a}$ is a row vector $(a_0,\, \dots, a_{\bar I})$ not collinear with $(1,\, \dots, 1)$ an $\bar I -1$ dimensional hyperplane $A \subset \Delta^0_{\bar I}$ is defined by $\mathbf{a}\,\tau$=0.
 
Note that in the case of $\bar{\mathcal{I}} \subset {\mathcal{I}} $  Assumption 1 in main text can be rephrased geometrically as:

 \begin{Assumption}
     \label{as:plane}
If $J\leq \bar I$ , $F^0$ is injective. 
If  $J > \bar I$ , then  any set   $ S=\{ F^0(p_{j_1}), \,\dots,\, F^0(p_{j_{\bar I + 1}})\} $  is affinely indipendent, namely it is not contained in a $\bar I-1$ hyperplane of  $\Delta^0_{\bar I} $.
 \end{Assumption}
  To get the geometric intuition for this see figure \ref{fig:plane}  in the case $\bar I = 2$
\begin{figure}[h!]
    \centering
    \caption{When $\bar I=2$ and $s=3$, the three points $a,b,c$ satisfy it and $x,y,w$ do not, but a small perturbation of, say, $x$ to $x'$ makes it satisfied.}
    \label{fig:plane}
\begin{tikzpicture}

\draw[-,ultra thick] (0,0)--(8,0) ;

\draw[-,ultra thick] (0,0)--(4,8) ;
\draw[-,ultra thick] (8,0)--(4,8) ;
    \coordinate (A) at (0,0);
    \coordinate (B) at (8,0);
    \coordinate (C) at (4,8);
       \fill[black!20, draw=black, thick] (A) circle (3pt) node[black, below left] {$0$};
              \fill[black!20, draw=black, thick] (B) circle (3pt) node[black, below right] {$Q_2$};
                  \fill[black!20, draw=black, thick] (C) circle (3pt) node[black, above right] {$Q_1$};

\coordinate (1) at (2.8,4.3);
   \fill[red, draw=red, thick] (1) circle (3pt)
   node[black, below left] {w};
   
\coordinate (1) at (3.8, 3.3);
   \fill[red, draw=red, thick] (1) circle (3pt)
   node[black, below left] {x};

\coordinate (1) at (3.9, 3.5);
   \fill[red!20, draw=red, thick] (1) circle (3pt)
   node[black, above right] {x'};

\coordinate (1) at (5.8, 1.3);
   \fill[red, draw=red, thick] (1) circle (3pt)
   node[black, below left] {y};
\draw[thin, dotted] (2.8, 4.3)--(5.8,1.3) ;

\coordinate (1) at (3.8,5.3);
   \fill[black, draw=black, thick] (1) circle (3pt)
   node[black, above right] {a};

\coordinate (1) at (5.1,4.3);
   \fill[black, draw=black, thick] (1) circle (3pt)
   node[black, above right] {b};

\coordinate (1) at (5.3,3);
   \fill[black, draw=black, thick] (1) circle (3pt)
   node[black, above right] {c};

\draw[thin, dotted] (3.8, 5.3)--(5.3,3) ;
    
\end{tikzpicture}
\end{figure}

Proof of the equivalence of the above Assumption and Assumption 1 in main text:
  A sequence of row operations transforms the matrix
\[
\big (\bar{F}_{j_{1}}, \dots, \bar{F}_{j_{\bar I + 1}} \big )= \begin{pmatrix}
1 & \dots & 1\\
F_{1,j_{1}} & \dots & F_{1,j_{\bar I + 1}}\\
F_{2,j_{1}} & \dots & F_{2,j_{\bar I + 1}}\\
& \dots & \\
\end{pmatrix}
   \] 
   into 
   \[ 
   F^0=\begin{pmatrix}
F_{0j_{1}}=1-\sum_{i \in  \bar I} F_{ij_{1}} & \dots & F_{0j_{\bar I + 1}}=1-\sum_{i \in  \bar I} F_{ij_{\bar I + 1}}\\
F_{1,j_{1}} & \dots & F_{1,j_{\bar I + 1}}\\
F_{2,j_{1}} & \dots & F_{2,j_{\bar I + 1}}\\
& \dots & \\

\end{pmatrix}
\]

Assumption  1 in the main text  requires these to be of full rank, this is equivalent to: for any nonzero $\mathbf{a}$ the equations $\mathbf{a} F^0 = 0 $ have no solution, i.e. that the columns  of  $F^0$ , the $F(p_{j_{l}})$,  are affinely independent. 
\qedsymbol

We introduce  the functions:
\begin{equation}
  L (\mathbf{\tau})= 1/b \sum_{i} b_i \,  ln \,  \tau_i \:,\, \tau \in \Delta_{ I-1}
\end{equation}
and, if $\bar {\mathcal{I}} \subset {\mathcal{I}}$: 

\begin{equation}
  L^0 (\mathbf{\tau})= 1/b \sum_{i\in \bar I} b_i \,  ln \,  \tau_i  \:,\, \tau \in \bar \Delta_{\bar I-1} or \, \; \Delta^0 _{\bar I} 
\end{equation}  so that $LL(\pi))= L (F(\pi))$ and, if $\bar {\mathcal{I}} \subset {\mathcal{I}}$,  
$LL(\pi))= L (F(\pi)) = L^0 (F^0(\pi)))$.

We summarize, spaces, maps and assumptions in them in the diagram:
\begin{equation}
\label{diag:LL}
  \begin{tikzcd}
 \Delta_{J-1} \arrow[r,hook,  "F"] \arrow[rr, "LL", bend left]
& \Delta_{I-1} \arrow[r, "L"] & \mathbf{R}
\end{tikzcd}   
\end{equation}
and 
\begin{equation}
\label{diag:LL^0}
 \begin{tikzcd}
 \Delta_{J-1} \arrow[drr, "LL" description , bend right= 60] \arrow[rd,tail ,  "F^0"] \arrow[r, "F"] & \Delta_{I-1} \arrow[d,two heads, "c"]&\\
& \Delta^0 \arrow[r, "L^0"] & \mathbf{R}  
\end{tikzcd}
\end{equation}
note that $L^0$ is a \textit{strictly} concave function defined on the  interior of  $\bar \Delta_{\bar I-1}$ and going to minus infinity on its boundary, 
it is  also \textit{strictly} concave on the interior of  $\Delta^0 _{\bar I}$  together with the interior of the front face $\bar \Delta_{\bar I-1}$, while it goes to minus infinity on the rest of its boundary.  So $LL$  is a concave function on  $\Delta_{J-1}$ and 
  if $ F$ is injective it is strictly concave in $\pi$. Note also that  $L ^0(((1-t), t\tau_1, \dots, t\tau_{\bar I}))= L(\mathbf{\tau}) + ln \, t $ and so it is a strictly increasing function of  $t$.

\section{Proofs}
\ {\sc \bf  Proof of  Theorem 1 in main text}

We first give the 

\begin{proof}[Proof of the Lemma in main text]
Let us define the function $g(\pi)= \sum_j \pi_j$. The first order conditions for a maximum of $LL$ with the constraint $g(\pi)=1$ ( i.e. on $\Delta(\pi)$ ) are:
\begin{eqnarray}
  \forall j, \;  (\nabla_j LL - \lambda \nabla_j g) \pi_j = 0 \\
\nabla_j LL - \lambda \nabla_j g \leq 0 , if \, \pi_j=0
\end{eqnarray}
and computing the derivatives we find that these conditions become:
\begin{equation}
 \label{eq:lambda}
  \forall j, \; 1/b \sum_i b_i (\frac{ F_{ij}}{\sum_k F_{ik}\pi_{k}} -\lambda ) \pi_j=0\\   
\end{equation}
\begin{equation}  
\label{eq:lambdas}
D_j(\pi)= 1/b \sum_i b_i (\frac{ F_{ij}}{\sum_k F_{ik}\pi_{k}} -\lambda) \leq 0, if \, \pi_j=0
\end{equation}
but its easy to see that $\lambda$ has to be equal to one: in fact summing equations  (main text, Ref.~X) over all the $j$ we get:
\begin{equation}
1= 1/b \sum_i b_i \frac{\sum_j F_{ij} \pi_j }{\sum_k F_{ik}\pi_{k}} = \lambda \sum_j \pi_j= \lambda 
\end{equation}

So equation \ref{eq:lambda} is exactly the fixed point equation required by the stability condition and equation  \ref{eq:lambdas}  is the condition for stability.  This ends the proof of the lemma.
\end{proof}

It follows that the stable fixed points are the maxima of $LL$, but $LL(\pi) = L (F (\pi) )$ and if $\bar {\mathcal{I}}={\mathcal{I}} $ then $L$ is strictly concave with a unique maximum in $\Delta_{I-1}$.  So in case of F injective the theorem is proved. 

Consider now the  case  $\bar {\mathcal{I}} \subset {\mathcal{I}}$ and  $J > \bar I $.
Consider the maps
\begin{equation}
\Delta_{J-1} \xrightarrow{ F^0} \Delta^0_{\bar I} \xrightarrow{L^0}  \mathbb{R}
\end{equation}

  Let $F^0(\Delta_{J-1})= K$, a compact, convex polytope.
If $J \leq \bar I < I $, Assumption  1 implies that  $F^0$ is injective.
If  $J > \bar I $  and $F$  satisfies  Assumption 1,  $F^0$ will be surjective as a linear map  and  $K$  will be of full dimension,  $\bar I$ , in $\Delta^0 $. In the interior of $K$ the map will in general not be injective if $J > \bar I + 1 $.  But let  $\partial K$ be its boundary in $\Delta^0$ and let $B= (F^0)^{-1} (\partial K)$ be its preimage.
 
 We have the following elementary result:
 
\begin{Lemma}
    \label{lm:geo}
    If Assumption is satisfied  the map $ F^0_{\mid B}: B \rightarrow \partial K$ is injective.
\end{Lemma} 
See Figure \ref{fig: geo} below for a geometric illustration.

\begin{figure}[h!]
    \centering
    \caption{ To illustrate Lemma 1: here $J=4$ , $\bar I =2$. Even if the preimage of interior points, $\tau_1$ may be a segment (green), on $\tau_2 \in \partial K$ (red), $F^0$ is injective.}
    \label{fig: geo}
\begin{tikzpicture}
\draw[-,ultra thick] (7,0)--(15,0) ;
\draw[-,ultra thick] (7,0)--(11,8) ;
\draw[-,ultra thick] (15,0)--(11,8) ;
    \coordinate (A) at (15,0);
    \coordinate (B) at (7.1,0);
    \coordinate (C) at (11,8);
\fill[black!20, draw=black, thick] (A) circle (3pt) node[black, below right] {$A$};
\fill[black!20, draw=black, thick] (B) circle (3pt) node[black, below left] {$B$};
\fill[black!20, draw=black, thick] (C) circle (3pt) node[black, above right] {$C$};
    
    \coordinate (a1) at (4,6);
    \coordinate (b1) at (4,2.8);
    \coordinate (d1) at (5.5,3.7);
    \coordinate (e1) at (1,4.2);

    \draw[line width=0.75mm, red, fill=black!10] (a1) -- (e1) ;
     \draw[thick, fill=black!10] (a1) -- (b1) ;
    \draw[line width=0.75mm,red, fill=black!20] (a1)  -- (d1) ;
 \draw[line width=0.75mm, red, fill=black!20] (b1)  -- (d1);
\draw[line width=0.75mm, red, fill=black!20] (b1)  -- (e1) ;
    \draw[thick, dash dot dot] (e1) -- (d1);

    \fill[black!20, draw=black, thick] (a1) circle (3pt) node[black, above right] {$b$};
    \fill[black!20, draw=black, thick] (b1) circle (3pt) node[black, below left] {$c$};

    \fill[black!20, draw=black, thick] (d1) circle (3pt) node[black, above right] {$d$};
    \fill[black!20, draw=black, thick] (e1) circle (3pt) node[black, below left] {$a$};
    
    \coordinate (a) at (11.8,4);
    \coordinate (b) at (10.8,0.8);
    \coordinate (d) at (13.1,1.2);
    \coordinate (e) at (8.8,3);

\coordinate (V3) at ($1.6*(a)-0.6*(d)$); 
\coordinate (V4) at ($-0.4*(a)+1.4*(d)$); 
\draw [dashed](V3)--(V4);
\coordinate (V5) at ($(V3)-(-0.2, 0.8)$); 
       \fill (V5) circle (0pt)
   node[black, above left] {$a \, \tau =0$};
   \coordinate (V6) at ($(V5)-(0.8, 1.2)$); 
       \fill (V6) circle (0pt)
   node[black, above left] {$ K $};

    \draw[line width=0.75mm, red, fill=black!10] (a) -- (e) -- cycle;
    \draw[line width=0.75mm, red, fill=black!20] (b) -- (d) -- cycle;
    \draw[line width=0.75mm, red, fill=black!20] (b) -- (e) -- cycle;
      \draw[line width=0.75mm, red, fill=black!20] (a) -- (d) -- cycle;
    \draw[thick, ] (e) -- (d);
     \draw[thick, fill=black!20] (a) -- (b);

    \fill[black!20, draw=black, thick] (a) circle (3pt) node[black, above right] {$b$};
    \fill[black!20, draw=black, thick] (b) circle (3pt) node[black, below left] {$c$};

    \fill[black!20, draw=black, thick] (d) circle (3pt) node[black, above right] {$d$};
    \fill[black!20, draw=black, thick] (e) circle (3pt) node[black, below left] {$a$};

\coordinate (K) at (10.5,1.9);
   \fill[green, draw=green, thick] (K) circle (3pt)
   node[black, above right] {$\tau_1$};
   
\coordinate (K) at (12.6,2.3);
   \fill[blue, draw=blue, thick] (K) circle (3pt)
   node[black, above right] {$\tau_2$};
   
\draw [->,>=stealth, thick] (5.5,4.5) -- (8,3.5);
    
\coordinate (K) at (4.8,4.7);
   \fill[blue, draw=blue, thick] (K) circle (3pt)
   node[black, above right] {$F_0^{-1}(\tau_2)$};
   
    \coordinate (j) at (3.1,4.9);
    \coordinate (h) at (3.3,3.6);
   \draw[thick,green] (j) -- (h);
\fill[green!20, draw=green, thick] (h) circle (3pt)
  node[black, above right] {$\pi_1$};
\fill[green!20, draw=green, thick] (j) circle (3pt)
  node[black, above right] {$\pi_1'$};
\coordinate (K) at (3.3,4);
   \fill[black!20, draw=black, thick] (K) circle (0pt)
   node[black, above left] {$F_0^{-1}(\tau_1)$};
   
    \coordinate (K) at (2.5,6.8);
   \fill[black!20, draw=black, thick] (K) circle (0pt)
   node[black, above right] {$\Delta_{J-1}$};
   
    \coordinate (K) at (13,6.8);
   \fill[black!20, draw=black, thick] (K) circle (0pt)
   node[black, above right] {$\Delta^0_{\bar I}$};

    \coordinate (K) at (6.5,4);
   \fill[black!20, draw=black, thick] (K) circle (0pt)
   node[black, above right] {$F_0$};
\end{tikzpicture}   
\end{figure}

 \begin{proof}[Proof of Lemma 1]
Let  $\tau_0 \in \partial K$ .  Suppose by contradiction that  it has  two distinct preimages  in 
$\Delta_{J-1}$ : $(\pi_1, \, \dots, \, \pi_J)\neq (\pi_1', \, \dots, \, \pi_J') $ and so   $\tau_0 = \sum \pi_i F(p_i)=\sum \pi_{i}' F(p_{i})$.
 By the separating hyperplane  theorem there exists an  hyperplane $A =\{ \tau \, | \, a \, \tau =0 \} $ such that $\mathbf{a}(\mathbf{\tau_0})=0$ and  $\mathbf{a}(\mathbf{\tau})>0$ if  $\mathbf{\tau}\in Int (K)$.  All the the points in  $S= \{F(p_i), \, i\,|\,\pi_i \, \text{or }\, \pi_i' \neq 0\} $  must lie on $A$: otherwise we would have say $\mathbf{a}(F(p_i))>0 $ and since $\alpha_i >0 $ and all other terms in the sum are $\geq 0$, we would have $\mathbf{a}(\tau_0)>0$. But then by Assumption 1  $S$  consists of at most  $\bar I$  points and these are  affinely independent, so the convex combination giving  $\mathbf{\tau_0}$ must be unique.
 \end{proof}
 
Since $K$ is compact and convex and  $L(\pi)$ is strictly concave on $\Delta^0 (\pi)$ it will have a unique maximum on some $ \mathbf{\tau_0} \in K$. We prove that this maximum cannot lie in the interior of $K$.
If $\mathbf{\tau_0}$ were in the interior it would be of the form: $ \mathbf{\tau_0}=((1-t_0),\, t_0\tau_1, \, \dots,\, t_0\tau_{\bar I})$ with  $0 < t < 1$  and,  for some $\epsilon >0 $ all points $ \mathbf{\tau}=((1-t), t\tau_1, \dots, t\tau_{\bar I});\; t \in [t_0, t_0 + \epsilon ]$ will also be in $K$, but as  $L$  is strictly increasing in t it cannot have a maximum there in  $\mathbf{\tau_0}$. So  $\mathbf{\tau_0}$ lies on $\partial K$, but because of the lemma it has a unique preimage, so $L(\pi)$ has a unique maximum in $\Delta(\pi)$.
\qedsymbol

\

\ {\sc \bf Proof of Theorem 2} The heuristic of the proof is to discretize the $x$ but we will proceed without appealing to it explicitly. Choose  $\delta_n >0,\;  n= 1,\, \dots, \, N $ .
Note that if we substitute $ \delta_n p(x_n \mid t_j)$ for    $p(x_n \mid t_j)$ the likelihood function changes by a constant: 
$$
L(\pi) =1/N \sum_n  \,  ln \, ( {\sum_k \delta_n p(x_n \mid t_k)\pi_{k}} )= 1/N  \sum_n ln \delta_n + 1/N \sum_n  \,  ln \, ( {\sum_k p(x_n \mid t_k)\pi_{k}} )
$$
so its maxima and its gradient are unchanged.
The same holds  for both the formulas for the fixed point:

\[
\pi_j =1/N \sum_n  \frac{ p(x_n \mid t_j)}{\sum_k p(x_n \mid t_k)\pi^0_{k}} \pi^0_j,  \ j=1,\dots,J.
\]
and the conditions for stability:
\begin{eqnarray}
  \forall j, \;  (\nabla_j L - \lambda \nabla_j g) \pi_j = 0 \\
\nabla_j L - \lambda \nabla_j g \leq 0 , if \, \pi_j=0
\end{eqnarray}
Choose now the  $\delta_n$  such that  $1- \sum_n \delta_n p(x_n \mid t_j) > 0, \, \forall j$ ,
and use the $\delta_n p(x_n \mid t_j)$ and $1- \sum_n \delta_n p(x_n \mid t_j) > 0$ to build a matrix such as the matrix $F=(F_{ij})$ in Section (main text, Ref.~X), it is immediate to see that Assumption 2 translates to Assumption 1 and so the proof is as in Theorem 1.
The matrices $\begin{pmatrix}  \delta_np(x_n \mid t_j) \end{pmatrix}$  and $\begin{pmatrix}  p(x_n \mid t_j) \end{pmatrix}$ have the same rank and also Assumption 1  remains valid after multiplication with $\delta_n$. \qedsymbol{}

\

{\sc \bf Proof of Theorem 3} Uniqueness follows as before, thanks to Assumption 3 that insures that the map on the boundary is injective. Maximization, too, follows as in Lemma 1 in the main text. As for convergence : suppose that a subsequence of $d\bar\pi_n$ converges weakly to some $d\tilde{\pi}$.  Weak continuity of $L$  implies that $L (F(d\bar\pi_n))$ converges to $L (F(d\tilde{\pi})) $.
On the other hand the density of $X_n$ in $X$ implies that $L(F(d\bar\pi_n))= \sup_ {\mathcal{P}(X_n)} L (F(d\pi))$ converges to 
$\sup_ {\mathcal{P}(X)} L\left(F(d\pi)\right)=L \left(F(d \bar\pi)\right )$, so $L (F(d \bar\pi))=L (F(d\tilde{\pi}))$  and by the uniqueness of the minimum on  $\mathcal{P}(X)$ , $d\bar{\pi}= d\tilde{\pi}$. Since this holds for any subsequence it follows that $d\bar \pi_n$ has $d\bar \pi$ as unique limit point. Note that we do not need to make any assumption on the $X_n$ and their maps: if there are several fixed points they all converge to the same $d\pi^0$.
\qedsymbol{} 

\

{\sc \bf Proof of Theorem 4} By passing to a subsequence  we can assume that  $d\bar\pi_k(t)$ converges  to $d\bar \pi(t)$. Let $d\pi(t)$ be any other measure in $\mathcal{P}(T)$. We set:
$ F\left( d\bar \pi_k\left(t\right)\right)= \bar f_k(x)dx$
$ F\left( d\bar \pi\left(t\right)\right)= \bar f(x)dx$
$ F\left( d\pi\left(t\right)\right)=  f(x)dx$. Let $d\mu_k$ be the empirical measures associated to the observations $x_1, x_2, \dots, x_k$; by the strong law of large numbers they converge weakly almost surely to $d\alpha$.%
\footnote{By the Glivenko-Cantelly theorem the convergence of is even uniform on distribution functions but we will not need it.}

By Theorem 3  $L \left(F(d \bar\pi_k)\right )=\int  ln  \bar f_k(x) \, d\mu_k \geq \int  ln  f(x) \, d\mu_k= L \left(F(d\pi)\right )$. The key point is to observe that since the $d\bar \pi_k\left(t\right)$ converge weakly and the operator $F$ is compact we have that $\lim_{k \rightarrow \infty }  \bar f_k(x)= \bar f(x)$  uniformly and so,  together with the weak convergence of the $d\bar \pi_k\left(t\right)$, this gives: 
$\int  ln  \bar f(x) \, d\alpha= \lim_{k \rightarrow \infty }  \int  ln  \bar f_k(x) \, d\mu_k$.
On the other side $\lim_{k \rightarrow \infty }\int  ln  f(x) \, d\mu_k= \int  ln  f(x) \, d\alpha$, all together this gives $\int  ln  \bar f(x) \, d\alpha \geq \int  ln  f(x) \, d\alpha$  
and substituting in the formula this gives us the result.
\qed

\end{document}